\documentclass{PoS}

\title{From p+p to Pb+Pb Collisions: \\ 
       Wounded Nucleon versus Statistical Models}

\ShortTitle{From p+p to Pb+Pb Collisions: Wounded Nucleon versus Statistical Models}

\author{\speaker{Marek Gazdzicki}%
        \\
        Goethe-University Frankfurt, Frankfurt am Main, Germany
        and \\ 
        Jan Kochanowski University, Kielce, Poland\\
        E-mail: \email{Marek.Gazdzicki@cern.ch}}

\abstract{
System size dependence of hadron production properties
is discussed within the Wounded Nucleon Model and 
the Statistical Model in the grand canonical, canonical
and micro-canonical formulations.
Similarities and differences between predictions of the
models related to the treatment of conservation laws are exposed.
A need for models which would combine
a hydrodynamical-like expansion with conservation laws obeyed
in individual collisions is stressed.

}

\FullConference{8th International Workshop on Critical Point and Onset of Deconfinement\\
                 March 11-15, 2013\\
                 Napa, California, USA}

\begin{document}

\section{Introduction}
\begin{figure}[t]
\centering
\includegraphics[width=0.5\textwidth]{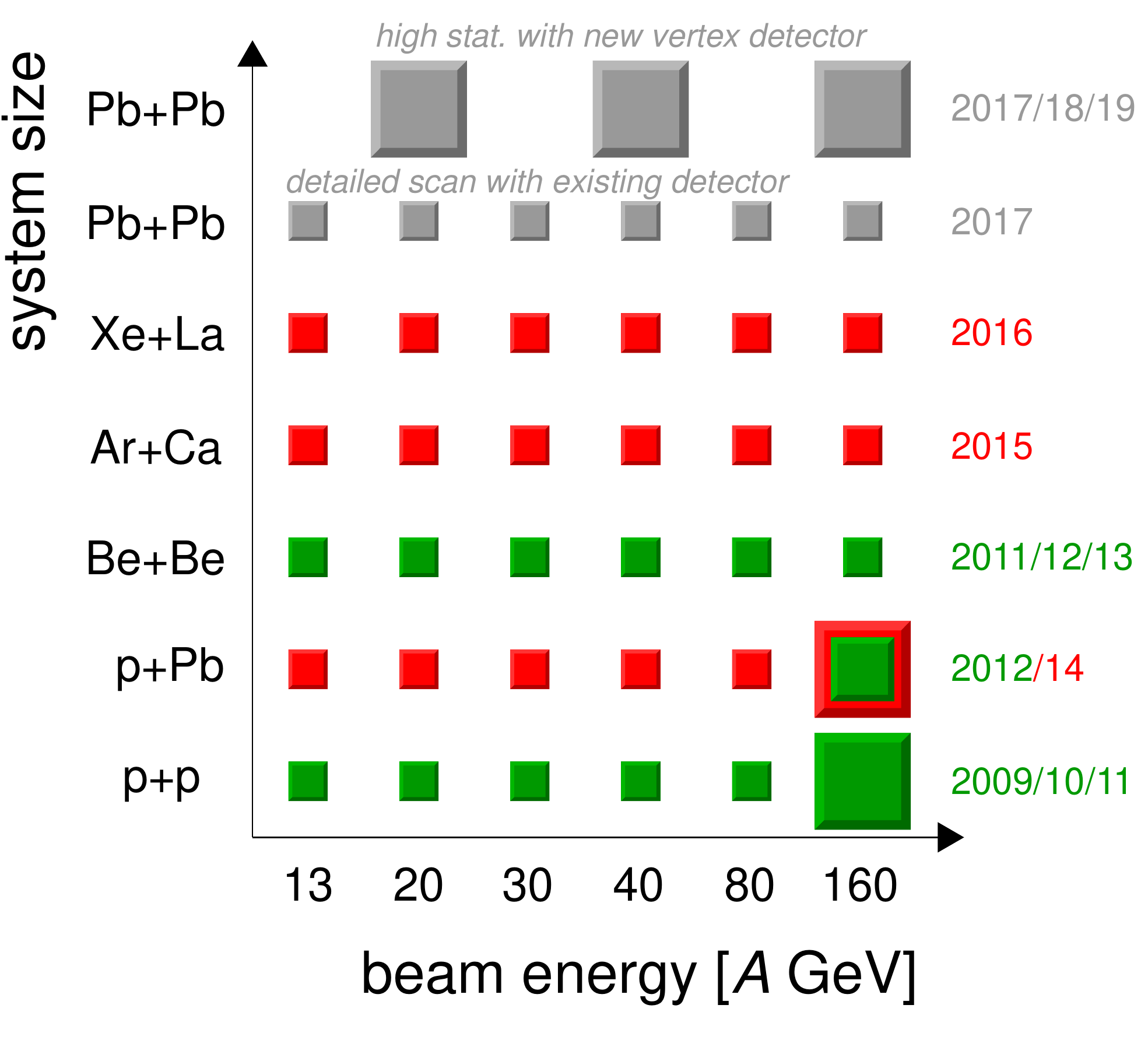}
\caption{
The NA61/SHINE data taking schedule for the ion program.
The already registered reactions are indicated by green squares, whereas
the approved future data taking and the proposed extension of 
the program are shown in red and gray, respectively. 
}
\label{fig:box_plot}
\end{figure}

This work is motivated by the NA61/SHINE ion 
program~\cite{na61proposal, cpod2013kasia}
in which hadron production properties are studied as a function
of size of colliding nuclei and their collision energy.
The status and plans of the NA61/SHINE data taking within this
program are presented in Fig.~\ref{fig:box_plot}.
The main goals of this program are the study of the onset
of deconfinement and the search for the critical point of strongly
interacting matter. They can be reached providing 
''trivial'' phenomena which affect the system size dependence 
and collision energy dependence are sufficiently understood.
In view of this I discuss here first the influence of 
fluctuations of the system size, and second 
the role of conservation laws on the system size dependence.
Two models are selected for this purpose:
the Wounded Nucleon Model~\cite{wnm} and 
the Statistical Model~\cite{fermi}.  
This is because they play a special role in physics
of heavy ion collisions, that is they are forefathers of the currently 
most popular approaches, the hydrodynamical and string-hadronic models.
Moreover, they continuously serve as the basic tools to interpret
results on hadron production in high energy collisions.
This is due to them being  simple and  approximately
reproducing several basic properties of the data.
They were formulated  before the QCD era and their relation
to QCD still remains  unclear (for a brief historical
review of multi-particle production in high energy collisions 
see Ref.~\cite{history}).

The paper is organized as follows.
First the models are introduced, then the similarities and
differences in their predictions concerning the system
size dependence are discussed. Closing remarks 
conclude the paper.

\section{The Wounded Nucleon Model}

\begin{figure}[t]
\centering
\includegraphics[width=0.5\textwidth]{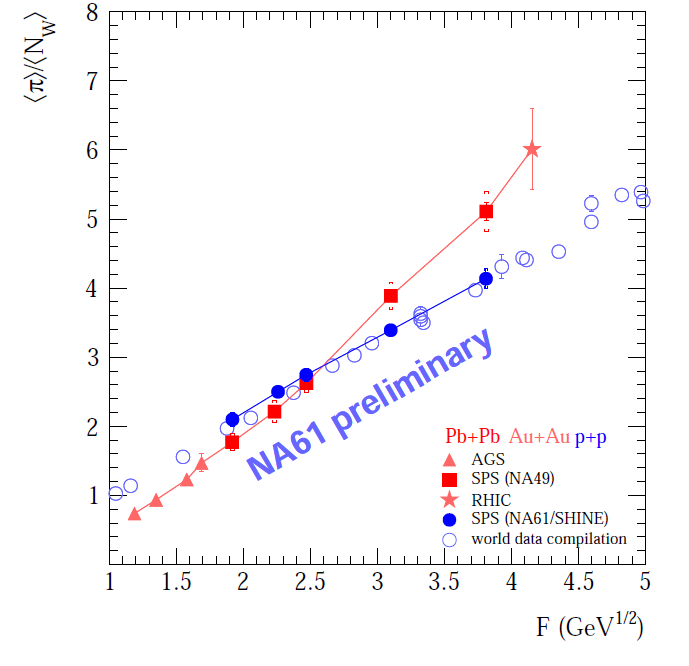}
\caption{
The ratio of mean pion multiplicity to the mean number of wounded nucleons
for p+p interactions and central Pb+Pb collisions
plotted as a function of collision energy~\cite{cpod_szymon} 
($F \approx s_{NN}^{1/4}$).
In the SPS energy region shown in the plot, the dependences for 
central Pb+Pb collisions and p+p interactions cross each other
at about 40$A$~GeV. At this energy the prediction of the WNM (see text)
is strictly valid.
}
\label{fig:pions}
\end{figure}

The Wounded Nucleon Model (WNM)
was proposed by Bialas, Bleszynski and Czyz~\cite{wnm} in 1976
as a late child of the S-matrix period~\cite{history}.
It assumes that particle production in nucleon-nucleon 
and nucleus-nucleus collisions is an incoherent superposition 
of particle production from wounded nucleons (nucleons 
which interacted inelastically and whose number is calculated using
the Glauber approach).
Properties of wounded nucleons are independent of the size of colliding
nuclei, e.g. they are the same in p+p and Pb+Pb collisions at the same
collision energy per nucleon.

The most famous prediction of the model reads:
\begin{equation}
\langle A \rangle / \langle W \rangle =  \langle A \rangle_{NN}/2~,
\label{eq:wnm}
\end{equation}
where $\langle A \rangle$ and $\langle A \rangle_{NN}$ is
the mean multiplicity of hadron $A$ in nucleus-nucleus collisions 
and nucleon-nucleon
interactions, respectively, whereas $\langle W \rangle$ and 2 are 
mean numbers
of wounded nucleons in nucleus-nucleus and nucleon-nucleon collisions.
This prediction is approximately valid for the most copiously produced hadrons,
pions. This is demonstrated in Fig.~\ref{fig:pions}, where 
the ratio of mean pion multiplicity to the mean number of wounded nucleons 
in p+p interactions and central Pb+Pb collisions
is plotted as a function of collision energy.
From the low SPS to the top RHIC (not shown in the plot) energies 
the ratio for p+p and Pb+Pb collisions differs 
''only'' by about 30\%~\cite{ood}.
This is why the WNM and its string-hadronic successors are still
in use.    

\section{The Statistical Model}

\begin{figure}[t]
\centering
\includegraphics[width=0.48\textwidth]{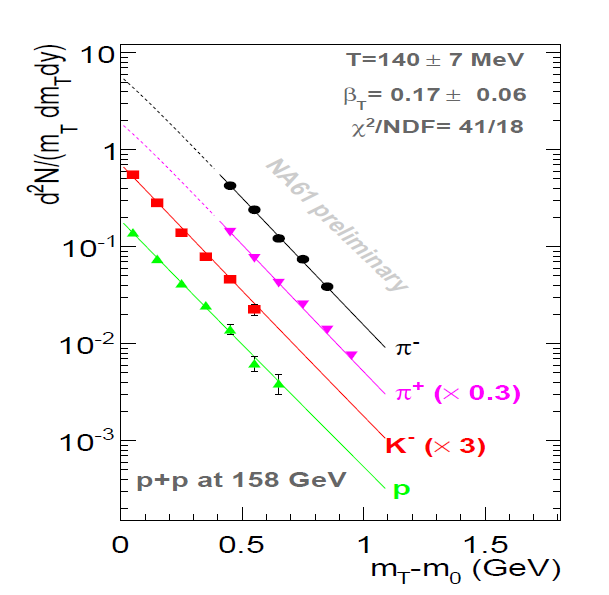}
\includegraphics[width=0.48\textwidth]{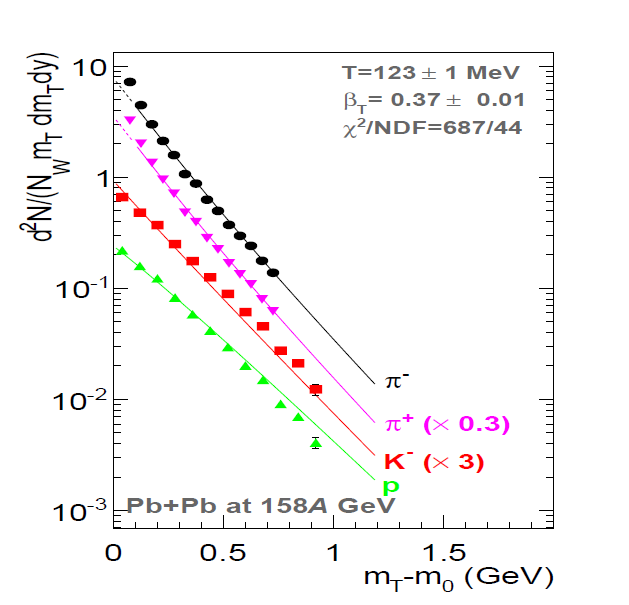}
\caption{
Examples of transverse mass spectra of the most abundant hadrons
produced in p+p interactions ($left$) and central
Pb+Pb collisions ($right$) at 158$A$~GeV~\cite{cpod_szymon}.
In p+p interactions the spectra are approximately exponential
with the inverse slope parameter, which is the same for different hadrons
as predicted by the Statistical Model.
The spectra in central Pb+Pb collisions deviate somewhat from
the prediction, which is interpreted as due to non-equilibrium
effects as the collective expansion of matter before freeze-out. 
}
\label{fig:mt}
\end{figure}
The Statistical Model of multi-particle production (SM) was initiated
by Fermi in 1950~\cite{fermi}.
Its basic assumption states that all possible micro-states of
the macroscopic system created in a collision are equally probable.
For large enough systems (e.g. central Pb+Pb collisions)
and abundantly produced hadron species (e.g. pions, kaons and protons) the 
grand canonical formulation of the model, SM(GCE), can be used for
particle inclusive spectra and mean multiplicities. This leads to 
the most famous prediction of the model, namely:
\begin{equation}
 d^3n/dp^3 \sim  V \cdot e^{-E/T} ~, 
\label{eq:stat}
\end{equation}
where $d^3n/dp^3$ is the particle momentum distribution and
$V$, $T$ and $E$ are system volume, its
temperature and particle energy, respectively.
This prediction agrees approximately with experimental data  if spectra 
in the direction transverse to the collision axis are considered.
Namely, distributions in the transverse mass of hadrons,
$m_T = \sqrt{p_T^2 + m^2}$, are measured to be approximately exponential for
$m_T \le 2$~GeV ($p_T$ and $m$ denote particle transverse 
momentum and mass, respectively).
This is shown in Fig.~\ref{fig:mt}, where 
examples of the transverse mass spectra in p+p interactions
($left$) and central Pb+Pb collisions ($right$) at 158$A$~GeV
are presented. 
This is why the SM(GCE) and its hydrodynamical successors
are in use today.

\section{System size dependence: Similarities}

In the WNM model hadrons are produced from ''decays'' of independent
wounded nucleons. Similarly in the SM(GCE) model they are
produced from ''decays'' of independent volume elements. 
This leads to almost identical predictions concerning the system size
dependence of hadron production properties in both models.
The detailed discussion of this conclusion is presented below
using as an example multiplicities
of two different types of hadrons, $A$ and $B$.
Two multiplicities are needed in order to express the model
predictions on the system size dependence so that they
are independent of the system size and its fluctuations.
It obviously simplifies a comparison between the models themselves 
as well as between the latter and experimental data.

First let us consider  mean multiplicities, 
$\langle A \rangle$ and $\langle B \rangle$.
In the WNM they are proportional to the number of wounded
nucleons:
\begin{equation}
 \langle A \rangle \sim W ~, ~~~~~~~~ 
 \langle B \rangle \sim W ~, 
\label{eq:wnm:mean}
\end{equation}
and in the SM(GCE) they are proportional to the system volume:
\begin{equation}
 \langle A \rangle \sim V ~, ~~~~~~~~
 \langle B \rangle \sim V ~. 
\label{eq:stat:mean}
\end{equation}
Obviously the ratio of mean multiplicities is independent of
the system size parameter, $W$ and $V$, in the WNM and the SM(GCE), respectively.
Moreover, it is easy to show that it is also independent of
the system size parameter fluctuations.
Thus, the ratio $\langle A \rangle \ / \langle B \rangle $ is
independent of $P(W)$ and $P(V)$, where $P(W)$ and $P(V)$ are
probability (density) distributions of $W$ and $V$, respectively,
for the considered set of collisions. 
Quantities which have the latter property are called strongly
intensive quantities~\cite{siq}. Such quantities should be used to 
study the system size dependence as they eliminate influence of
usually poorly known distributions of the system size parameters,
$W$ and $V$. 

Second we consider multiplicity fluctuations characterized
by second moments of multiplicity distributions.
Two quantities of relevance are variance,
$Var[X] = \langle (X - \langle X \rangle)^2 \rangle$ and
scaled variance, $\omega[X] = Var[X]/\langle X \rangle$,
where $X$ stands for $A$ or $B$. 

The scaled variances of $A$ and $B$ and the mixed second
moment $\langle AB \rangle$ calculated within the WNM read~\cite{siq}:
\begin{equation}
 \omega[A] = \omega^*[A] + \langle A \rangle / \langle W \rangle \cdot \omega[W] ~, \\ 
\label{eq:wnm:varA}
\end{equation}
\begin{equation}
 \omega[B] = \omega^*[B] + \langle B \rangle / \langle W \rangle \cdot \omega[W] ~, \\
\label{eq:wnm:varB}
\end{equation}
\begin{equation}
 \langle AB \rangle = \langle AB \rangle^ * \langle W \rangle +
 \langle A \rangle  \langle B \rangle \langle W \rangle^2 \cdot
 (\langle W^2 \rangle - \langle W \rangle )~.
\label{eq:wnm:AB}
\end{equation}
As may be expected they have a similar form in the SM(GCE)~\cite{siq}:
\begin{equation}
 \omega[A] = \omega^*[A] + \langle A \rangle / \langle V \rangle \cdot\omega[V] ~, \\ 
\label{eq:stat:varA}
\end{equation}
\begin{equation}
 \omega[B] = \omega^*[B] + \langle B \rangle / \langle V \rangle \cdot\omega[V] ~, \\
\label{eq:stat:varB}
\end{equation}
\begin{equation}
 \langle AB \rangle = \langle AB \rangle^ * \langle V \rangle +
 \langle A \rangle  \langle B \rangle \langle V \rangle^2 \cdot
 (\langle V^2 \rangle - \langle V \rangle )~.
\label{eq:stat:AB}
\end{equation}
In Eqs.~\ref{eq:wnm:varA}-\ref{eq:stat:AB} quantities denoted by $^*$
are quantities calculated for any fixed value of the system size
parameter, $W$ and $V$, within the WNM and the SM(GCE), respectively.   

From Eqs.~\ref{eq:wnm:varA}-\ref{eq:wnm:AB} and
Eqs.~\ref{eq:stat:varA}-\ref{eq:stat:AB} follows~\cite{siq,sign} that properly
constructed functions of the second moments, namely

\begin{equation}
 \Delta[A,B] = ( \langle B \rangle \omega[A] -
                 \langle A \rangle \omega[B] )~/~
               ( \langle B \rangle - \langle A \rangle )  
\label{eq:delta}
\end{equation}
and
\begin{equation}
 \Sigma[A,B] = ( \langle B \rangle \omega[A] +
                 \langle A \rangle \omega[B] - 
                 2 (\langle AB \rangle - \langle A \rangle \langle B \rangle) )~/~
               ( \langle B \rangle + \langle A \rangle )  
\label{eq:sigma}
\end{equation}
are independent of $P(W)$ and $P(V)$ in the WNM and the SM(GCE), respectively.
Thus $\Delta[A,B]$ and $\Sigma[A,B]$ are strongly intensive quantities
which measure fluctuations, i.e. they are sensitive to second moments of
$A$ and $B$  distributions.
The $\Sigma$ quantity is a reincarnation of the popular
$\phi$ measure of fluctuations~\cite{phi}.
As the quantities $\Delta[A,B]$, $\Sigma[A,B]$ and $\phi$ are
strongly intensive their values in p+p and Pb+Pb collisions are equal.
Of course, this is valid only within the WNM and the SM(GCE).
At this conference preliminary experimental results on the $\phi$
quantity in p+p and central Pb+Pb collisions at the CERN SPS energies
were presented~\cite{cpod_maja}. They concern three choices of 
$A$ and $B$ hadron multiplicities, namely [$A$, $B$] = [$K$, $\pi$],
[$p$, $\pi$] and [$K$, $p$], where $K$, $\pi$ and $p$ denote multiplicities
of charged kaons, pions and protons, respectively.
Noticeably, the results for p+p interactions are close to the ones
for central Pb+Pb collisions.

In summary, predictions of the WNM and the SM(GCE) concerning the system
size dependence are very similar. In particular, the models predict that
the mean multiplicity ratio as well as
the $\Delta$,  $\Sigma$ and $\phi$ quantities are independent 
of the system size and its fluctuations.
The number of wounded nucleons - the system size parameter in the WNM - 
is discreet, whereas
the volume - the system size parameter in the SM(GCE) - is continous. 
This may lead to somewhat different predictions
for quantities which are dependent on the system size parameter
distribution.

\begin{figure}[t]
\centering
\includegraphics[width=0.48\textwidth]{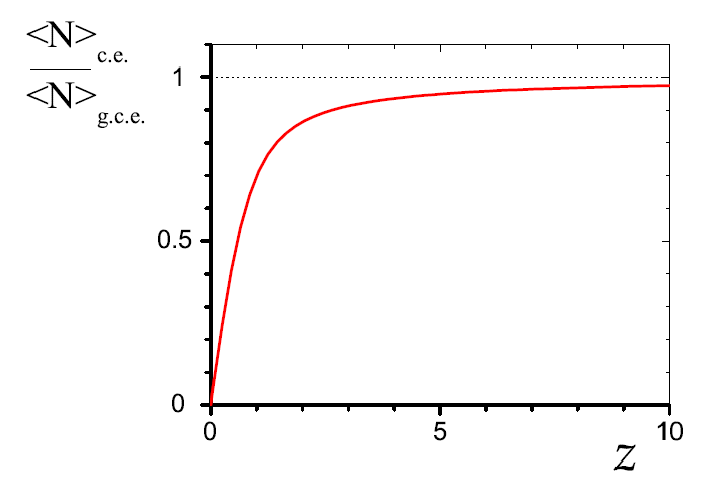}
\includegraphics[width=0.48\textwidth]{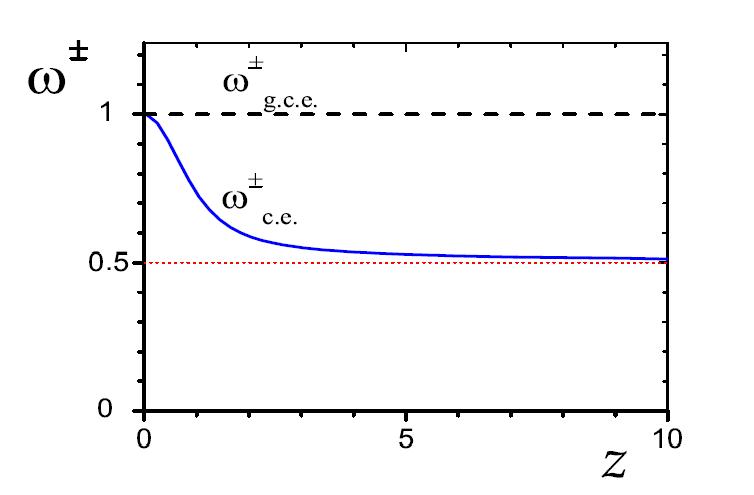}
\caption{
The ratio of  mean multiplicities  
($left$) and scaled variances ($right$)
calculated within the SM(CE) and the SM(GCE) are 
plotted as a function of  mean multiplicity $z$ for the SM(GCE), the latter 
being proportional to the system volume.
The ideal gas model of classical positively and negatively
charged particles is used for calculations.
}
\label{fig:ideal}
\end{figure}

\section{System size dependence: Differences}

Predictions of the Statistical Model concerning the volume
dependence change qualitatively when material and/or motional
conservation laws are introduced, i.e. instead of the grand
canonical ensemble, the canonical (CE) or micro-canonical (MCE)
ensembles are used.
The effect of conservation laws has been extensively studied
for mean multiplicities since 
1980~(see e.g. Refs.~\cite{rd,tr,bf}) and
for second moments of multiplicity distributions since 
2004~(see e.g. Refs.~\cite{begun_ce,begun_mce}). 
Three examples are presented below to illustrate the main results.

Figure~\ref{fig:ideal} 
taken from Ref.~\cite{begun_ce}
presents the results of calculations
performed within the simplest model which allows to study
the effect of the material conservation laws on mean multiplicity and
scaled variance of the multiplicity distribution.
In the model the ideal gas of classical positively and negatively 
charged particles is assumed.
The ratio of the mean multiplicities calculated within the SM(CE) 
and the SM(GCE) is
plotted in Fig.~\ref{fig:ideal} ($left$)
as a function of the mean multiplicity from the SM(GCE), 
the latter being proportional
to the system volume.
The ratio approaches one with increasing volume. 
Thus for sufficiently large systems mean multiplicities obtained 
within the SM(GCE) can be used instead of mean multiplicities
from the SM(CE) and the SM(MCE)~\cite{begun_mce}.
This is however not the case for the scaled variance as illustrated
in Fig.~\ref{fig:ideal} ($right$).
The results for the SM(CE) and the SM(GCE) approach each other when the volume
decreases to zero. Of course, the scaled variance in the SM(GCE) is  one
independent of volume. 
Different behaviour is observed for the scaled variance in the SM(CE), it 
decreases with increasing volume
and for a sufficiently large volume  approaches 0.5.

\begin{figure}[t]
\centering
\includegraphics[width=0.50\textwidth]{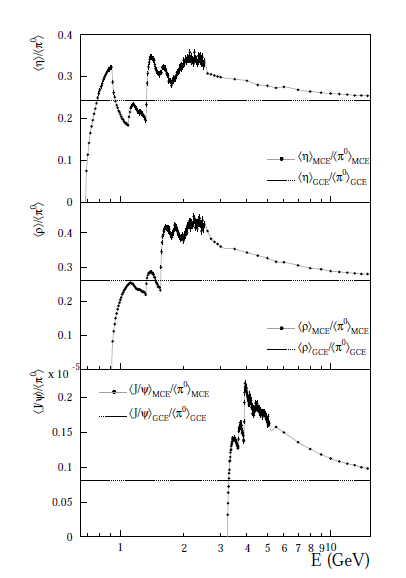}
\caption{
Ratios of mean multiplicities of neutral hadrons calculated within the MCE and GCE 
formulations of the hadron-resonance gas model are
plotted as a function of the system energy, the latter being proportional
to the system volume in the GCE.
}
\label{fig:mean_hrg}
\end{figure}

Finally, the comparison of mean hadron multiplicities calculated
with the GCE and MCE formulations of the hadron resonance gas model~\cite{fe_mce}
is shown in Fig.~\ref{fig:mean_hrg}.   
For small energies of the system (in the GCE energy is proportional to volume)
the multiplicity ratios strongly depend on system energy and volume.
The observed peaks and dips are correlated with the thresholds for
the production of various hadrons.

\section{Closing remarks} 

\begin{enumerate}
\item
I demonstrated that the predictions of the Wounded
Nucleon Model and the Statistical Model in the grand canonical
formulation concerning the system size dependence 
are almost identical. Differences may appear only because
the system size parameter is discreet in the WNM
(the number of wounded nucleons) and continuous in the SM(GCE)
(the system volume).
The similarity of predictions explain why these models and their
successors (string-hadronic and hydrodynamical models)
can co-exist.
\item
We all believe that material and motional conservation laws
are obeyed in each high energy collision.
Within the Statistical Model their influence can be studied
by a comparison of the model predictions obtained using different
ensembles, GCE, CE and MCE. 
The result is that the predictions are strongly dependent 
on conservation laws. For small systems mostly mean multiplicities
are affected whereas for large systems scaled variances
are modified.
Thus the GCE formulation of the Statistical Model is
neither valid for small nor for large systems 
when both mean multiplicities and fluctuations are of relevance.
\item
Rich experimental data on hadron spectra in heavy ion 
collisions favor hydrodynamical models.
In these models, similarly to the SM(GCE), 
conservation laws are obeyed only for
averages over many collisions.
This significantly limits their applicability.
Thus, it seems to be urgent to
develop models which would
combine a hydrodynamical-like expansion with conservation
laws obeyed in individual collisions.
Fortunately this effort has already started~\cite{hydro}.
\end{enumerate}

\vspace{0.3 cm}
\begin{acknowledgments}  I am thankful to Mark Gorenstein and Peter Seyboth
for comments and corrections. This work was
supported by 
German Research
Foundation (grants DFG GA 1480/2-1 and GA 1480/2-2).
\end{acknowledgments}

\end{document}